\documentclass[twocolumn,aps,english,prl,preprintnumbers]{revtex4}  
\usepackage{graphicx}  
\usepackage{amssymb}   
\usepackage{amsmath}

\usepackage{babel}
\usepackage{color}
\usepackage{slashed}
\usepackage{mathrsfs}

\begin{document}





\title{Dark Matter from a Dark Connection}

\author{Ranit Das$^{a}$}
\email{ranit.das@stonybrook.edu}
\author{Chethan Krishnan$^{b}$}
\email{chethan.krishnan@gmail.com}

\affiliation{$^{a}$Physics and Astronomy Department, Stony Brook University, Stony Brook, NY 11794-3800, \\}
\affiliation{$^{b}$Center for High Energy Physics,  Indian Institute of Science, Bangalore 560012\\}

\date{\today}

\begin{abstract}

In the first part of this note, we observe that a  non-Riemannian piece in the affine connection (a ``dark connection") leads to an algebraically determined, conserved, symmetric 2-tensor in the Einstein field equations that is a natural dark matter candidate. The only other effect it has, is through its coupling to standard model fermions via covariant derivatives. If the local dark matter density is the result of a background classical dark connection, these Yukawa-like mass corrections are minuscule ($\sim 10^{-31}$ eV for terrestrial fermions) and {\em none} of the tests of general relativity or the equivalence principle are affected. In the second part of the note, we give dynamics to the dark connection and show how it can be re-interpreted in terms of conventional dark matter particles. The simplest way to do this is to treat it as a composite field involving scalars or vectors. The (pseudo-)scalar model naturally has a perturbative shift-symmetry and leads to versions of the Fuzzy Dark Matter (FDM) scenario that has recently become popular (eg., arXiv:1610.08297) as an alternative to WIMPs. A vector model with a ${\cal Z}_2$-parity falls into the Planckian Interacting Dark Matter (PIDM) paradigm, introduced in arXiv:1511.03278. It is possible to construct versions of these theories that yield the correct relic density, fit with inflation, and are falsifiable in the next round of CMB experiments. Our work is an explicit demonstration that the meaningful distinction is not between gravity modification and dark matter, but between theories with extra fields and those without.





\end{abstract}

\maketitle


\section{Introduction}

In this article, first, we show that a non-Riemannian piece in the affine connection can  be indistinguishable from dark matter-energy. The age-old approach for discussing connection dynamics is that of Palatini, but there is a natural alternative, exploiting the fact that an arbitrary connection can be written as the sum of the Levi-Civita connection and a tensor field that we call the dark connection \footnote{The usual Palatini (or related) approach forces us to describe the full dynamics of the connection from the get go. Here we have the possibility of treating the dark connection as an effective background field, whose dynamics we can deal with separately if we feel up to the task.}. This structure leads to a covariantly conserved symmetric 2-tensor in the field equations that is algebraically determined in terms of the dark connection, with the natural interpretation as a dark matter-energy stress tensor \footnote{Note that despite the many distinct pieces of compelling evidence that we have for the existence of dark matter at various scales, so far, {\em all} of this evidence is reliant on its gravitational coupling. In other words, we only know dark matter-energy at the level of a (phenomenological) stress tensor, we do not know what is it that constitutes this stress tensor.}. 

If we treat the dark connection as a background classical field (constrained only by conservation laws), we show that the fluid stress tensors relevant for large scale structure (including $\Lambda$CDM) can be reverse-engineered from it. Because it contains fields other than the metric, this approach can evade the standard criticisms against classes of gravity modification that may run into trouble with Bullet cluster \cite{bullet} or ``dark-matter-less" galaxies \cite{dmlessgalaxy}. 

Standard model fermions are affected by covariant derivatives, so if the local dark matter density in the Solar System ($\sim 6 \times 10^{-22}$ kg/m$^3$) is due to a classical dark connection, it leads to a mass correction of $\Delta m \sim 10^{-31}$ eV to fermionic fields.  This is somewhat analogous to how fermions acquire Yukawa mass terms via the vacuum expectation value of Higgs. We argue that such a dark connection background is not in conflict with {\em any} of the experimental tests of general relativity or the principle of equivalence \footnote{We might be tempted to make the glib claim that modifying the connection is merely the introduction of a tensor field, which can (presumably) be interpreted as yet another form matter. But unlike a typical matter field, a piece in the connection can affect both covariant derivatives and affine geodesics, so this is a little too quick. In other words, conventional forms of matter are coupled to gravity via covariantization with the Levi-Civita connection, whereas here, the covariantization itself is modified. So one needs to make sure that  fifth force effects, equivalence principle violations, etc. are within bounds. }.

In the second half of this note, we consider ways to give dynamics to the dark connection. We show how simple models can reproduce conventional dark matter scenarios. 
The simplest approaches treat the dark connection as a  composite field involving scalars or vectors. The (pseudo-)scalar model  has a natural shift symmetry and leads to Fuzzy Dark Matter (FDM) which can be a scalar or an axion-like particle. This is viable: it was considered recently in eg., \cite{WittenFDM} and follow-ups, as an approach that circumvents various difficulties with the WIMP/CDM picture. If instead the dark connection is made of a (massive) vector particle, it can be related to the Planckian Interacting Dark Matter (PIDM) paradigm \cite{PIDM1, PIDM2}. If we declare that the dark connection has a discrete reflection symmetry, it becomes identical to the vector models considered in \cite{PIDM2}. This is again a viable scenario.  





Our goal in this article is to investigate one of the most natural versions of gravity modification \footnote{The need for dark matter to explain the IR physics of the Universe can be viewed as the need for modifying the Newtonian potential at long length scales. In Einstein's gravity, the gravitational potential has a natural interpretation as the connection, so it stands to reason that modifying the physics of the connection can be an interesting thing to try. Note that MOND \cite{Milgrom, teves} can in fact be interpreted as a modification of the Newtonian potential via introducing a new scale. We find this way of viewing MOND, namely as a modification of {\em field theory} rather than as a modification of {\em dynamics}, more palatable.}. Remarkably, we find that in explicit examples, it ends up being closely related to viable dark {\em matter} models. We view this as part of a more general statement. When there are fields other than the metric that go  into the description of gravity, it is somewhat arbitrary what one calls matter and what one calls gravity.  
The real distinction therefore is not between modified gravity and dark matter models, but between theories with extra fields and those without them. In other words, the challenge for a puritanical gravity modifier is to explain dark matter phenomenology {\em without} adding extra fields. We note in passing that covariant (partially successful) gravity modification attempts like TeVeS \cite{teves}, contains many extra fields and is consistent with this expectation.

Details suppressed in this short note will appear in a follow-up paper \cite{followup}.




\section{Dark Matter from a Dark Connection}

We will take the spacetime connection to be  
${\Gamma_{bc}}^a= \Big\{ ^a_{bc} \Big\} + {\Delta_{bc}}^a$
where $\Big\{ ^a_{bc} \Big\}$ is the standard Levi-Civita connection and $\Delta^{a}_{bc}$ is a tensor, that is symmetric in the lower indices. This tensor field, we call the {\em dark connection}. This above structure captures the most general (symmetric) connection one is allowed to have on a manifold with a metric \footnote{We could add a torsion if we want, but it will not be necessary to make our point. Our interest is in symmetric modifications of the connection. A very general class of such theories has recently been presented in \cite{CK}.}. Usually in general relativity, $\Delta$ is assumed to vanish. Instead we consider the Einstein-Hilbert action with the modified connection:
\begin{equation}\label{3} S
=\int \left[{1 \over 2\kappa }\,R (g,\{\}+\Delta)+{\mathcal  {L}}_{{\mathrm  {SM}}}(g, \Delta, \psi)\right]{\sqrt  {-g}}\,{\mathrm  {d}}^{4}x \end{equation}
with $\kappa=8\pi G $, $c=1$, and $R(g,\Gamma)=g^{ab}R_{ab}(\Gamma)$. ${\mathcal  {L}}_{{\mathrm  {SM}}}$ is the matter (standard model) action which can depend on $\Delta$ through covariant derivatives. $\psi$ stands for all the standard model fields.
Note the crucial but elementary fact that the Riemann and Ricci tensors do not depend on the metric, only on the connection. Note also that we have not specified the dynamics of $\Delta$ at the moment, 
we will give it dynamics via some explicit models later. 

By direct calculation it turns out that
$R_{ab}(\Gamma)= R_{ab} (\{ \}) + {\tilde R}_{ab}(\Delta)$ 
where the $R_{ab}(\{ \})$ is the standard Ricci tensor computed with the Levi-Civita connection, and 
${\tilde R}_{a b}(\Delta)\equiv \nabla^{LC}_{c}{{\Delta}_{b a}}^c-\nabla^{LC}_{b} {\Delta_{c a}}^c+{\Delta_{c d}}^c {\Delta_{b a}}^d-{\Delta_{b d}}^c {\Delta_{c a}}^d$. 
$\nabla^{LC}$ is the covariant derivative with the Levi-Civita connection. The first two terms in this expression enter the Lagrangian as 
$\sqrt{-g} g^{ab}(\nabla^{LC}_{c}{{\Delta}_{b a}}^c-\nabla^{LC}_{b} {\Delta_{c a}}^c)$ 
which turns out to be a total derivative. Upon ignoring the boundary terms that arise from them, our action is 
\begin{equation}\label{8} S=\int \left[{1 \over 2\kappa }\,g^{ab}\Big(R_{ab}(\{\})+ {\mathcal{D}}_{ab} \Big)+
{\mathcal  {L}}_{{\mathrm  {SM}}}(g,\Delta,\psi)\right]{\sqrt  {-g}}\,{\mathrm  {d}}^{4}x \end{equation}
The extra piece in the Lagrangian is algebraically \footnote{The explicit expressions we write below assume that the dark connection has no explicit dependence on the metric. If we treat the dark connection as a composite field with algebraic dependence on the metric,  the explicit expression for the dark connection stress tensor  will change, but our qualitative statements remain unaffected \cite{followup}. We will consider such composite field examples, when we discuss dynamics.} determined in terms of the dark connection:
\begin{equation}\label{9}{\mathcal{D}}_{ab} = {\Delta_{c d}}^c {\Delta_{b a}}^d-{\Delta_{b d}}^c {\Delta_{c a}}^d 
\end{equation}
The equation of motion for the metric is
\begin{equation}\label{10}R_{{a b }}(\{\})-{\frac  {1}{2}}g_{{a b }}R(\{\}) =\kappa (T_{{a b }}+\mathcal{T}_{ab}). \end{equation}
This is the usual Einstein field equation, with the usual matter stress tensor
$T_{{a b }}={\frac  {-2}{{\sqrt  {-g}}}}{\frac  {\delta ({\sqrt  {-g}}{\mathcal  {L}}_{{\mathrm  {SM}}})}{\delta g^{{a b }}}}$,
except that now we also have a new term 
\begin{equation}\label{12}
\mathcal{T}_{ab}=\frac{1}{\kappa}\left(\frac{1}{2}g_{ab}g^{cd}{\mathcal{D}}_{cd}-{\mathcal{D}}_{ab}\right).
\end{equation}
It is symmetric in its indices, and is covariantly conserved. This makes it a good candidate for a dark sector stress tensor \footnote{Note that positivity conditions of the stress tensor expectation value on a general state are not typically derivable from the matter Lagrangian directly, except with simplifying assumptions.}. Crucially, it is under the Levi-Civita connection (and {\em not} the full connection) that the dark connection stress tensor is covariantly conserved  \footnote{This fact can be used to argue that whenever stress tensor distributions are supported on curves in spacetime, those curves are forced to be {\em metric} geodesics, as required for the equivalence principle to hold for macroscopic test particles.  The basic idea behind this requires a bit of elaboration and goes back to two results, one due to Papapetrou \cite{Pap-Dixon} and the other to a theorem of Geroch and Jang \cite{Geroch}. We will present the details elsewhere; in this note we will give alternative arguments that show that equivalence principle violations are within bounds.}. This follows from the diff invariance of the theory, and because the stress tensor arises from the metric equation of motion.





\section{Dark Connection as a Background}

\subsection{Reverse Engineering Stress Tensors}

A fluid stress tensor (with generally, a time-dependent equation of state) is the most general symmetric tensor that respects the isometries of the FRW background \footnote{A similar discussion can be made for other backgrounds with other isometries (say Schwarzschild) as well, if one's goal is to model galactic dark matter \cite{followup}.}. 
Applying the same idea to the dark connection and writing the result in a general (not necessarily co-moving) frame, we find
\begin{equation}\label{16}
{\Delta_{bc}}^a= \alpha\ U^a g^{FRW}_{bc}+\beta\ (U_c \delta^a_b+ U_b \delta^a_c)+ \gamma\ U^a U_b U_c \end{equation}
where $\alpha$, $\beta$ and $\gamma$ can in general be time dependent. 
With $g_{ab}U^{a}U^{b}=-1$, and using the definition of the dark connection stress tensor \eqref{9}\&\eqref{12}, we have  \begin{eqnarray}\label{17} 
\mathcal{T}_{ab}= \frac{1}{2} \left(-\alpha ^2+\alpha  (\gamma -6 \beta )+3 \beta  (\gamma -\beta )\right) g^{FRW}_{ab} + \hspace{0.3in}\nonumber \\ 
+ (\alpha ^2-\alpha  \gamma -3 \beta ^2+3 \beta  \gamma) U_{a}U_{b}. \hspace{0.1in}
\end{eqnarray}
Note that this has precisely the form of a perfect fluid 
$T_{ab}=p g_{ab} +(p+\rho)U_a U_b$. 
Generically, we would expect that one can reproduce an arbitrary perfect fluid stress tensor from a dark connection using the above expression. Indeed, the regions of real $\alpha$-$\beta$-$\gamma$-space that can reproduce $\Lambda$CDM, pressureless dust, radiation, and specific values of the equation of state $w$, are easily charted out, even though we will not present the details here \footnote{See \cite{followup}. Just to emphasize that not anything goes, note that if we were to set one of $\alpha, \beta, \gamma$ to zero, many interesting cases are ruled out.}. This serves as a sanity check. We will aim to model only dark matter (or pressureless dust).



\subsection{Fermionic Mass Corrections}

Apart from the contribution to the stress tensor that arises from the Einstein-Hilbert piece, the dark connection couples to standard model fields as well, via covariant derivatives. The correct coupling to gauge fields, and in particular light, is essential for our theory to not die at birth. Happily,  the connection term drops out trivially from the coupling of gauge fields and Higgs to gravity \footnote{For the Higgs, and scalars in general, this depends on whether one writes the kinetic term as $ \phi \Box \phi$ or $\partial \phi \partial \phi$, by adjusting boundary terms. For the moment, we are assuming the latter form. If we don't, a version of the discussion for fermions will apply to the scalars as well. We thank M. Headrick for a comment on this.}, so it is only the fermionic sector of the standard model that we have to worry about. For fermions, the presence of the dark connection will show up as a non-standard spin connection. If we decide to view the dark matter stress tensor as due to a background dark connection, this will imply that the dark connection piece will be analogous \footnote{Analogous, but not identical. This is for two reasons. Firstly, the dark connection field is not  in the vacuum state, it has spatial and temporal variations because dark matter has dynamics on large scales. Secondly, the term arises from a piece in the covariant derivative, not from an actual Yukawa term.} to a vev in a Yukawa term and will lead to mass corrections to the fermions. This is worrisome, because this would indicate a small violation of the principle of equivalence.

To see whether this is serious, let us estimate the value of this mass correction \footnote{The precise details of the fermionic terms are straightforward to write down \cite{followup}, but we will not need anything more than some judicious dimensional analysis to make our claims in this paper.}. Since the dark matter stress tensor is dust and since its dependence on the dark connection is quadratic (see previous section), we have upon re-instating the dimensionful constants:
\begin{eqnarray}
 \sqrt{G_N \rho}\sim  \langle \Delta_{bc}^{\ \ a} \rangle \sim \frac{\delta m \  c^2}{\hbar}  
\end{eqnarray}
The ``vev" of $\Delta$ leads to the mass correction of the fermion, and the $\hbar$ arises for the same reason that it arises in the Klein-Gordon mass, see eqn (1.21) in \cite{Srednicki}, or the discussion near the end of section II in \cite{WittenFDM}. These expressions essentially follow from dimensional analysis and the structure of the terms in the Lagrangian: the $G_N$ can appear only via the Einstein-Hilbert piece, therefore the rest of the dimensions have to be soaked up with $\hbar$'s and $c$'s. Now, using the fact that the dark matter density in the solar system is $\rho \sim 6 \times 10^{-22}$ kg/m$^3$, we find  that the mass correction is
\begin{eqnarray}
\delta m \sim 10^{-31} {\rm eV}.
\end{eqnarray}
This is an incredibly small value, just two orders of magnitude away from the current Hubble scale. To put some perspective, neutrino masses are expected to be of the order of an eV, and so will be corrected at the $10^{-31}$ level. An electron's mass will be corrected by a fraction of $10^{-39}$. This means that to capture this with a macroscopic experiment (even ignoring the fact that most of the mass is from QCD effects) one will need a resolution of $\lesssim 10^{-39}$. The Eot-Wash experiments that have the best bounds on Equivalence principle have only a $10^{-13}$ resolution, and the best Lorentz violation bounds are at the $10^{-22}$ level \cite{EotWash}. In other words, interpreting the local dark matter density as a classical background dark connection is entirely consistent with current experiments \footnote{There is a different class of equivalence principle experiments that have to do with measuring the (mostly time) variation of fundamental constants involving non-gravitational forces, like the weak force strength between protons \cite{EotWash}. Since dark matter scales like dust with the scale factor, the dark connection as well as the mass correction $\delta m$ must scale as $a(t)^{-3/2} \sim 1/t$ during the matter dominated era, where $t$ is the age of the Universe. Using these, the correction to fermion masses a (few) billion or so years ago can be estimated to be again not substantially different from $\sim 10^{-31}$ eV. So this is again perfectly consistent with experimental bounds, which are from relatively small red shifts ($\sim$ age of the Earth). If one views the success of Big Bang Nucleosynthesis as a test of the principle of equivalence (as done in Table 1 of \cite{EotWash}), then that provides us with a test from the early Universe, and we can no longer do the very generic analysis we have done here: we need specific models. The models we consider in the next section (thanks to the fact that they turn out to fall into phenomenologically viable classes) provide precisely that.}. 

\subsection{A Dynamical Classical Field?}

Viewing the dark connection as a classical background (analogous to the Higgs vev or quintessence) is tempting, but it also raises some challenges:
\begin{itemize}
\item Unlike the Higgs, the time dependence on cosmological scales is vital here \footnote{As well as spatial dependence, if one wants to understand structure formation. But presumably, the spatial dependence can be understood as classical growth of perturbations around a homogeneous background, as is usually done in cosmology.}. So one cannot just minimize a potential (say), one has to worry about the full dynamics \footnote{This is similar to the k-essence/quintessence idea, where one treats dark matter/energy as a dynamical scalar field of some type.}. 
\item This is further complicated by the fact that the dark connection is not a scalar, so either one has to think about it as a (massive) higher spin field or as an effective/composite field, when giving it dynamics. 
\end{itemize}
In the rest of this note, we will explore the possibility of treating the dark connection as a composite field \footnote{The idea of treating it as a higher spin field is also intriguing, and since the background is highly symmetric, we strongly suspect that some progress can be made. However, a consistent theory involving (elementary) higher spins requires an infinite tower of them and a framework like string theory, so we will not explore it further here.}.  We will see that the simplest models for the dark connection, lead readily to viable dark matter candidates. In fact, we will find that one of our models leads to ultra light (pseudo-)scalars, which indeed realize dark matter via coherent oscillations.



\section{Dark Connection as Dark Matter}

Since the dark connection term comes with a factor of Newton's constant, the resulting physics is naturally at the Planck scale: our dark matter is not a WIMP. In specific models, we will see (a) an ``axion" decay constant near the Planck scale resulting in a Fuzzy Dark Matter (FDM) candidate, and (b) a stable Planckian Interacting Dark Matter (PIDM) relic with (near-)Planckian mass that is decoupled from the standard model. The discussion below is self-contained enough to show that the models we end up with are subsets of previously known viable models, more details of the phenomenology will be presented in \cite{followup}.

\subsection{Fuzzy Dark Matter (FDM)}

The simplest possibility is to set
\begin{eqnarray}
\Delta^a_{bc} = \mu \ g_{bc} \partial^a \phi 
\end{eqnarray}
where $\phi$ is a (pseudo-)scalar and $\mu$ is a dimensionless parameter we have introduced for later convenience. Note that for consistency, $\phi$ has to be dimensionless, and the presence of the derivative immediately suggest that it should be viewed as a (pseudo-) Nambu-goldstone boson (pNGB) with a perturbative shift symmetry. Note that the shift symmetry is our key interest here, as it was in \cite{WittenFDM}. This is required for forbidding perturbative generation of Planck scale masses. Whether the field is truly an axion, in the sense that it is a pseudo-scalar pseudo-Goldstone boson of a broken chiral symmetry, will not be too important for our purposes \footnote{More generally, the phenomenology of a pNGB is controlled by two scales: the scale of spontaneous symmetry breaking $F$ (essentially the Planck scale for us), and the scale of explicit breaking $\nu$, which is controlled by non-perturbative effects. The mass of the pNGB is $m^2 \sim \nu^4/F$, and its couplings are suppressed by $1/F$. In this paper, following \cite{WittenFDM} we will assume that $\nu$ is such that the mass of the pNGB is in the FDM regime. Most discussion of pNGBs for dark matter purposes are in the context of axion like particles, but see \cite{Jaeckl} for a somewhat more general discussion. More general pNGBs have also appeared in the context of composite scalar dark matter models, see eg., \cite{Pomarol}.}. Axionic fields arise naturally in stringy set ups, and generic particles of this type with couplings to $F \tilde F$ terms are referred to as axion-like particles (ALP) in the astro-particle community \footnote{See, eg., \cite{Ranjan} for more references on axion like particles and searches for them, but in a different energy range.}. There exist plausible mechanisms to non-perturbatively generate  masses in the $10^{-22}$ eV range \cite{WittenFDM} (as required for a viable FDM candidate). We will further assume that the field $\phi$ is periodic \cite{WittenFDM}, to ensure that the non-perturbative breaking of shift symmetry leads to the usual sinusoidal potential. 


Substituting the above form and expanding our previous action, we find that indeed we have a kinetic term for a massless particle with ``axion" decay constant $F\sim \mu M_{Pl}$. The only other relevant terms are the (drastically suppressed) derivative couplings to SM fermions with strength $1/F$, see eg., footnote 5 in \cite{WittenFDM}. This suppression is a related, but distinct avatar of the statement we made in the previous section that the couplings of the standard model fermions to a background dark connection are  harmless. However, we are told that even though standard searches do not work, there are constraints via gravitational interactions, which deserve more scrutiny \footnote{We thank R. Laha for this comment.}.

Notably, models of this type have recently received an immense amount of attention, see \cite{WittenFDM} and follow-ups. Three broad  motivations behind these non-WIMP models are: (1) the failure of direct detection attempts so far, (2) the failure so far to detect TeV-scale SUSY at the LHC, and (3) various problems in explaining physics at the sub-galactic scale using cold dark matter. For $M_{Pl} \gtrsim F \gtrsim  10^{-2} M_{Pl}$, it was shown that a coherently oscillating field might solve these problems while retaining the successes of $\Lambda$CDM \footnote{This involves some plausible assumptions about the parameter controlling the non-perturbatively generated potential that breaks the shift symmetry \cite{WittenFDM}.}. Furthermore there exist mechanisms for adequately producing them cosmologically, see \cite{Marsh}. For $1 \gtrsim \mu \gtrsim 10^{-2}$,  our model falls into this paradigm. 

\subsection{Planckian Interacting Dark Matter (PIDM)}

Ultra-light scalars share some of the features of the classical field approach we mentioned in the previous section, because they are after all coherently oscillating configurations. However it is also possible to construct dark connection-inspired models where we end up with massive relic {\em particles}. A simple way to realize this is to set
\begin{eqnarray}
\Delta^a_{bc} = \mu\ g_{bc} A^a
\end {eqnarray}
where $A^a$ is a vector field. This term in the action leads to a Planck-scale mass term (suppressed by factors of $\mu$) for the vector field, and to make it dynamical we can add a kinetic term. This turns out to be a specific realization of the PIDM scenario of {\cite{PIDM1} with a massive vector as the dark matter particle \cite{PIDM2}.

We need to ensure that the SM fermionic couplings to the vector field are removed in some way, so that the Planck mass particle can be a stable relic. A simple way to ensure this, is to declare a ${\cal Z}_2$ parity symmetry for the action under which $A^a \rightarrow -A^a$. The SM couplings (and only those) drop out when we demand this. In \cite{PIDM1, PIDM2} an unspecified global symmetry was invoked to rule out the dark sector-SM sector direct coupling, the ${\cal Z}_2$-parity is our instantiation of this \footnote{Global symmetries are violated by non-perturbative effects in quantum gravity, this raises the possibility of direct detection, see \cite{PIDM2, followup}.}. Another plausible way to ensure this is to view the massive vector as arising suitably from the Higgsing of a dark sector gauge group, under which all the visible sector particles are singlets.

The dark connection particle naturally has a mass around the Planck scale in this set up (as well as in the PIDM paradigm in general). But one can have a closer to GUT scale mass if we allow $\lesssim 1\%$ fine-tuning. In our case, this translates to $\mu \lesssim 10^{-2}$. Such a relic can be produced with the right abundance during reheating after inflation via the freeze-in mechanism \cite{freeze-in, PIDM1, PIDM2}, see also related previous discussions in the context of WIMPZILLAs \cite{wz1, wz2}. For higher masses, the reheating temperature has to be high to produce it adequately, and that runs into tension with bounds on tensor-to-scalar ratio. Note that the model is falsifiable: it predicts a high enough tensor-to-scalar ratio in the CMB power spectrum to be detectable in the next round of observations \cite{PIDM1}. A more detailed discussion of the phenomenology is straightforwardly adapted from \cite{PIDM2}, but we will not present it here \cite{followup}.

\section{Discussion}
We conclude by summarizing our main takeaways:
\begin{itemize}
\item A non-Riemannian piece in the affine connection translates to a candidate for a dark matter(-energy) stress tensor.
\vspace{-0.1in}
\item If the dark connection were a classical field, it would be consistent with all local experimental tests while producing the needed dark matter component in the stress tensor.
\vspace{-0.1in}
\item A simple way to give dynamics to the dark connection and to end up with conventional models is to treat it as a composite field.
This can lead to particle physics models for dark matter that are fully phenomenologically viable.
\vspace{-0.1in}
\item Surprisingly (at least to us!), the simplest models that emerge this way have previously been considered for independent reasons. In particular, the (pseudo-)scalar models lead to FDM and the vector model is closely related to a PIDM. 
\vspace{-0.1in}
\item We find it notable that the simplest composite dark connection model can be re-interpreted as an ultra light particle (protected by a perturbative shift symmetry) with Planckian-suppressed couplings with the SM. When axionic, such fields are ubiquitous in string theory \cite{Svrcek}.
\vspace{-0.1in}
\item The dark connection is an explicit demonstration that the meaningful distinction is not between gravity modification and dark matter, but between theories with and without extra fields.
\end{itemize}


\vspace{0.2in}
We thank Parameswaran Ajith, Daniele S. M. Alves, Matteo Baggioli, Pallab Basu,  Jyotirmoy Bhattacharya, Biplob Bhattacherjee, Simon Carron-Huot, Shiuli Chatterjee, Oleg Evnin, Jarah Evslin, Dimitrios Giataganas, Matt Headrick,  Prasad Hegde, Ranjan Laha, Suvrat Raju, Bindusar Sahoo, Xerxes Tata and Sudhir Vempati for discusssions on various closely related matters. We are grateful to Biplob Bhattacherjee for pointers to literature on uses of pNGBs, and to Ranjan Laha for comments on a previous version of the manuscript. CK thanks the Vietnam Academy of Science and Technology (VAST) and the Aspen Center for Physics (supported by National Science Foundation grant PHY-1607611) for hospitality during some of this work. CK's work was partially supported by a grant from the Simons Foundation.


\begin{thebibliography}{99}


\bibitem{bullet}
https://en.wikipedia.org/wiki/Bullet$\_$Cluster

\bibitem{dmlessgalaxy} Pieter van Dokkum, et al. ``A galaxy lacking dark matter",  	Nature, 555, 629 (2018).

\bibitem{WittenFDM} 
  L.~Hui, J.~P.~Ostriker, S.~Tremaine and E.~Witten,
  ``Ultralight scalars as cosmological dark matter,''
  Phys.\ Rev.\ D {\bf 95}, no. 4, 043541 (2017)
  doi:10.1103/PhysRevD.95.043541
  [arXiv:1610.08297 [astro-ph.CO]].

\bibitem{PIDM1} 
  M.~Garny, M.~Sandora and M.~S.~Sloth,
  ``Planckian Interacting Massive Particles as Dark Matter,''
  Phys.\ Rev.\ Lett.\  {\bf 116}, no. 10, 101302 (2016)
  doi:10.1103/PhysRevLett.116.101302
  [arXiv:1511.03278 [hep-ph]].

\bibitem{PIDM2} 
  M.~Garny, A.~Palessandro, M.~Sandora and M.~S.~Sloth,
  ``Theory and Phenomenology of Planckian Interacting Massive Particles as Dark Matter,''
  JCAP {\bf 1802}, no. 02, 027 (2018)
  doi:10.1088/1475-7516/2018/02/027
  [arXiv:1709.09688 [hep-ph]].
  

\bibitem{followup} R. Das and C. Krishnan, ``Phenomenology of the Dark Connection", to appear.

\bibitem{CK} 
  C.~Krishnan,
  ``A Generalization of Gravity,''
  Found.\ Phys.\  {\bf 45}, no. 12, 1574 (2015)
  doi:10.1007/s10701-015-9941-2
  [arXiv:1409.6757 [hep-th]].


\bibitem{Milgrom} 
M.~Milgrom,
  ``A Modification of the Newtonian dynamics as a possible alternative to the hidden mass hypothesis,''
  Astrophys.\ J.\  {\bf 270}, 365 (1983).
  doi:10.1086/161130


\bibitem{teves} 
  J.~D.~Bekenstein,
  ``Relativistic gravitation theory for the MOND paradigm,''
  Phys.\ Rev.\ D {\bf 70}, 083509 (2004)
  Erratum: [Phys.\ Rev.\ D {\bf 71}, 069901 (2005)]
  doi:10.1103/PhysRevD.70.083509, 10.1103/PhysRevD.71.069901
  [astro-ph/0403694].





\bibitem{Pap-Dixon} 
  A.~Papapetrou,
  ``Spinning test particles in general relativity. 1.,''
  Proc.\ Roy.\ Soc.\ Lond.\ A {\bf 209}, 248 (1951).
  doi:10.1098/rspa.1951.0200



\bibitem{Geroch}
R.~P.~Geroch and P.~S.~Jang,
  ``Motion of a body in general relativity,''
  J.\ Math.\ Phys.\  {\bf 16}, 65 (1975).
  doi:10.1063/1.522416

\bibitem{Srednicki} 
  M.~Srednicki,
  ``Quantum field theory,'' Cambridge 2007.




\bibitem{EotWash}
  C.~M.~Will,
  ``The Confrontation between General Relativity and Experiment,''
  Living Rev.\ Rel.\  {\bf 17}, 4 (2014)
  doi:10.12942/lrr-2014-4
  [arXiv:1403.7377 [gr-qc]].

\bibitem{Jaeckl} 
  J.~Jaeckel, V.~M.~Mehta and L.~T.~Witkowski,
  ```Monodromy Dark Matter,''
  JCAP {\bf 1701}, no. 01, 036 (2017)
  doi:10.1088/1475-7516/2017/01/036
  [arXiv:1605.01367 [hep-ph]].

\bibitem{Pomarol} 
  M.~Frigerio, A.~Pomarol, F.~Riva and A.~Urbano,
  ``Composite Scalar Dark Matter,''
  JHEP {\bf 1207}, 015 (2012)
  doi:10.1007/JHEP07(2012)015
  [arXiv:1204.2808 [hep-ph]].

\bibitem{Ranjan} 
  H.~Vogel, R.~Laha and M.~Meyer,
  ``Diffuse axion-like particle searches,''
  arXiv:1712.01839 [hep-ph].


\bibitem{Marsh} 
  A.~Diez-Tejedor and D.~J.~E.~Marsh,
  ``Cosmological production of ultralight dark matter axions,''
  arXiv:1702.02116 [hep-ph].
  
\bibitem{freeze-in} 
  L.~J.~Hall, K.~Jedamzik, J.~March-Russell and S.~M.~West,
  ``Freeze-In Production of FIMP Dark Matter,''
  JHEP {\bf 1003}, 080 (2010)
  doi:10.1007/JHEP03(2010)080
  [arXiv:0911.1120 [hep-ph]].

\bibitem{wz1} 
  E.~W.~Kolb, D.~J.~H.~Chung and A.~Riotto,
  ``WIMPzillas!,''
  AIP Conf.\ Proc.\  {\bf 484}, no. 1, 91 (1999)
  doi:10.1063/1.59655
  [hep-ph/9810361].

\bibitem{wz2} 
  G.~F.~Giudice, E.~W.~Kolb and A.~Riotto,
  ``Largest temperature of the radiation era and its cosmological implications,''
  Phys.\ Rev.\ D {\bf 64}, 023508 (2001)
  doi:10.1103/PhysRevD.64.023508
  [hep-ph/0005123].

\bibitem{Svrcek} 
  P.~Svrcek and E.~Witten,
  ``Axions In String Theory,''
  JHEP {\bf 0606}, 051 (2006)
  doi:10.1088/1126-6708/2006/06/051
  [hep-th/0605206].
  
\end{thebibliography}
\end{document}